\documentclass[12pt,superscripaddress]{iopart}
\usepackage{iopams}  
\usepackage{graphicx}
\usepackage{epsfig, epstopdf}
\usepackage{bm}
\newcommand {\Fig}[1] {Figure~\ref{#1}}

\newcommand{\beq}{\begin{equation}}
\newcommand{\eeq}{\end{equation}}

\newcommand{\natc}{N@C$_{60}$} 
\newcommand{\natcseventy}{N@C$_{70}$}

\newcommand{\patc}{P@C$_{60}$}

\newcommand{\csixty}{C$_{60}$}
\newcommand{\cseventy}{C$_{70}$}

\newcommand{\cstwo}{CS$_2$}

\newcommand{\nfourteen}{$^{14}$N}
\newcommand{\nfifteen}{$^{15}$N}
\newcommand{\cthirteen}{$^{13}$C}

\newcommand{\beqa}{\begin{eqnarray}}
\newcommand{\eeqa}{\end{eqnarray}}
\newcommand{\w}{\omega}

\newcommand{\ACIE}{Ang. Chem. Int. Ed.}
\newcommand{\APL}{Appl. Phys. Lett.}

\newcommand{\CPL}{Chem. Phys. Lett.}

\newcommand{\JACS}{J. Am. Chem. Soc.}

\newcommand{\MOLPHYS}{Mol. Phys.}

\newcommand{\PTRSA}{Phil.~Trans.~R.~Soc. A}
\newcommand{\PRA}{Phys. Rev. A}
\newcommand{\PRB}{Phys. Rev. B}

\begin{document}

\title{Towards a fullerene-based quantum computer}
 
\author{Simon~C~Benjamin$^{1}$, Arzhang~Ardavan$^2$, G~Andrew~D~Briggs$^1$, David A Britz$^1$, Daniel Gunlycke$^1$, John Jefferson$^3$, Mark~A~G~Jones$^1$, David F Leigh$^1$, Brendon W Lovett$^1$, Andrei~N~Khlobystov$^{1,4}$, S~A~Lyon$^5$, John~J~L~Morton$^{1,2}$, Kyriakos~Porfyrakis$^1$, Mark~R~Sambrook$^1$, Alexei M Tyryshkin$^5$}

\address{$^1$ Department of Materials, University of Oxford, Parks Rd., Oxford, OX1 3PH, UK}
\address{$^2$ Clarendon Laboratory, University of Oxford, Parks Rd., Oxford, OX1 3PU, UK}
\address{$^3$ QinetiQ, St. Andrews Road, Malvern, WR14 3PS, UK}
\address{$^4$ School of Chemistry, University of Nottingham, University Park, Nottingham, NG7 2RD, UK}
\address{$^5$ Department of Electrical Engineering, Princeton University, Princeton, NJ 08544, USA}
\ead{simon.benjamin@materials.ox.ac.uk}

\date{\today}

\begin{abstract}
Molecular structures appear to be natural candidates for a quantum technology: individual atoms can support quantum superpositions for long periods, and such atoms can in principle be embedded in a permanent molecular scaffolding to form an array. This would be true nanotechnology, with dimensions of order of a nanometre. However, the challenges of realising such a vision are immense. One must identify a suitable elementary unit and demonstrate its merits for qubit storage and manipulation, including input / output. These units must then be formed into large arrays corresponding to an functional quantum architecture, including a mechanism for gate operations. Here we report our efforts, both experimental and theoretical, to create such a technology based on endohedral fullerenes or `buckyballs'. We describe our successes with respect to these criteria, along with the obstacles we are currently facing and the questions that remain to be addressed. 

\end{abstract}

\pacs{76.30.-v, 81.05.Tp}

\maketitle

\section{Introduction}

\begin{figure}
\centerline {\includegraphics[width=5.5in]{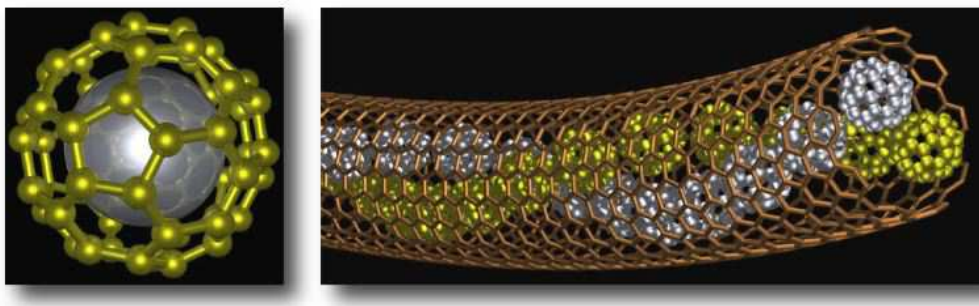}}
\caption{Left: A model of \natc, illustrating that the nitrogen atom sits at the centre of
the fullerene cage. Its electron wavefunction lies almost entirely inside, extending on the
cage with only a 2\% overlap. Right: An illustration of one of the fullerene array structures we have created: the `peapod' nanotube contains fullerenes packed in a pseudo-helical phase.} \label{nc60cartoon2}
 \end{figure}

In the quantum information processing (QIP) field there is a sense of optimism that a matter-based quantum computer can be built. That a scalable solid state implementation will be feasible remains to be demonstrated. Eventual quantum computers may well require a hierarchy of embodiments of quantum information, with weakly interacting stationary qubits for relatively long term storage, more strongly interacting partially delocalised qubits for controlled gates, and weakly interacting propagating qubits for communication. Within this hierarchy, electron spins appear to offer versatile properties, with reasonably long coherence times and the potential for interactions of controllable strength. By choosing materials in which the spin-orbit interaction is small, it is possible to modulate spatial distributions, and hence interactions, without otherwise affecting the quantum information stored in the spin. Carbon nanotubes offer one-dimensional electronic structures with low spin-orbit coupling and a choice of electronic structures, and endohedral fullerenes offer almost perfectly isolated atomic properties incarcerated in a carbon cage (illustrated in \Fig{nc60cartoon2})~\cite{harneit, briggsRS}. The chemical properties of these nanomaterials allow molecular assembly to complement lithography to create designer nanostructures with atomic precision in structure and reproducibility. The remarkable progress that has been made in ion trap quantum computing has been attributed in part to the deep understanding of the properties of the components and their interactions. The more we learn about the properties of carbon nanomaterials for quantum computing, the more promising they seem to be. In this paper we report progress in developing our vision of exploiting endohedral fullerenes and nanotubes for QIP.

We begin by describing the elementary fullerene unit, and the properties which we have established for qubit storage and manipulation. We describe efforts to optically address these units as an alternate means of manipulation and moreover as a mechanism for information input and output. We outline the research we have undertaken to create arrays of these fullerene units, both at the level of few qubit systems (e.g. dimers) and extended, scalable structures. We then discuss the theoretical issues and opportunities presented by the qubit-qubit interactions in such structures. Finally, we remark on the prospects for two-dimensional (or higher) molecular arrays as a fully fledged quantum computer technology. 
 
\section{An endohedral fullerene qubit}
\subsection{The \natc~spin system}

\begin{figure}[t]
\centerline {\includegraphics[width=6in]{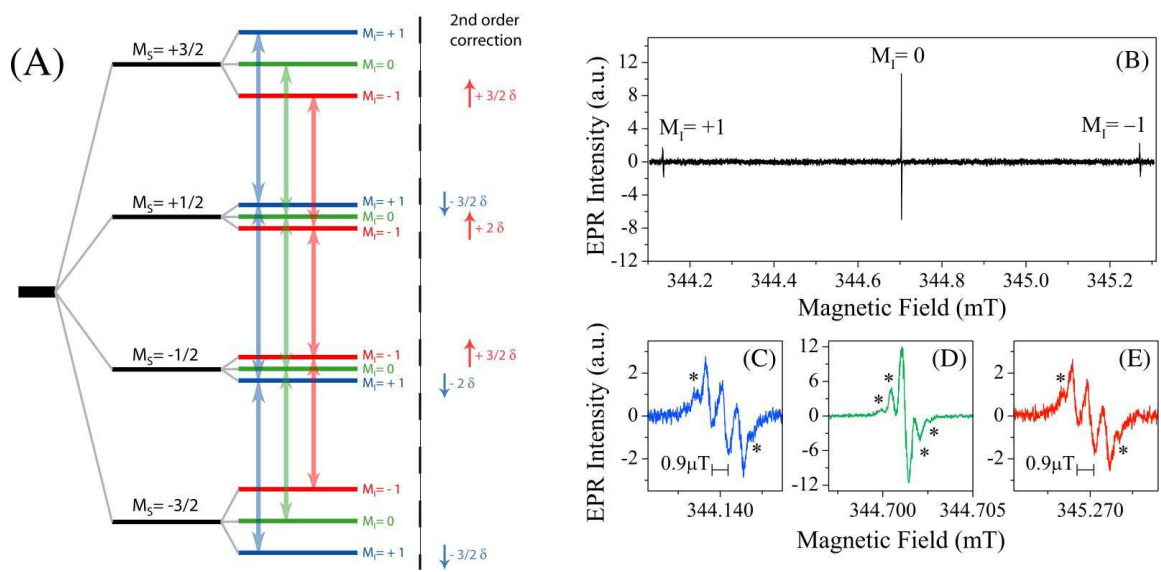}} \caption{(A) $^{14}$\natc~has electron spin $S=3/2$ and nuclear spin $I=1$ which together provide a rich 12 level structure. Considering only the first order hyperfine interaction, the three electron transitions associated with a particular nuclear spin projection are degenerate. Adding the second-order corrections ($\delta=a^2/B$), lifts the degeneracies for the $M_I=\pm1$ lines. (B) Continuous wave EPR spectrum of high purity N@C$_{60}$ in degassed CS$_{2}$ at room temperature. Each line in the triplet signal is labeled with the corresponding projection $M_I$ of the $^{14}$N nuclear spin. (C-E) Zoom-in for each of the three hyperfine lines reveals further structure. Stars (*) mark the line split by 13C hyperfine interactions with C60 cage. Measurement parameters: microwave frequency, 9.67~GHz; microwave power, 0.5~$\mu$W; modulation amplitude, 2~mG; modulation frequency, 1.6~kHz. .}\label{cwEPR}
\end{figure}

The molecule \natc~(that is, a nitrogen atom in a C$_{60}$ cage, for which the IUPAC notation is \emph{i}-NC$_{60}$) has electron spin $S=3/2$ coupled to the $^{14}$N nuclear spin
$I=1$ via an isotropic hyperfine interaction. The spin
Hamiltonian is therefore
\begin{equation}\label{Hamiltonian}
\mathcal{H}_0=\w_e S_z + \w_I I_z + a \vec{S} \!\cdot\! \vec{I},
\end{equation}
where $\w_e=g\beta B_0/\hbar$ and $\w_I=g_I\beta_n B_0/\hbar$ are
the electron and $^{14}$N nuclear Zeeman frequencies, $g$ and $g_I$
are the electron and nuclear g-factors, $\beta$ and $\beta_n$ are
the Bohr and nuclear magnetons, $\hbar$ is Planck's constant and
$B_0$ is the magnetic field applied along $z$-axis in the laboratory
frame. This Hamiltonian yields the 12-level system illustrated in
\Fig{cwEPR}. The electron spin resonance frequency is
primarily determined by the electron Zeeman term; this is then
further split by a hyperfine interaction with the $^{14}$N nucleus.

A continuous-wave EPR spectrum of
\natc ~at room temperature prepared using established methods~\cite{morton04} is shown~in \cstwo. The spectrum is
centered on the electron g-factor $g=2.0036$ and comprises three narrow
lines (linewidth $<0.3~\mu$T) resulting from the hyperfine coupling to $^{14}$N
\cite{Murphy1996}. The three hyperfine lines
are not of equal amplitude, the outer two being approximately 30$\%$
of the height of the central line. This can be attributed to
broadening of the outer two lines with respect to the central $I=0$
line~---~ a result of \emph{second order} hyperfine splitting. The
full hyperfine term is
\beq 
a \vec{S} \!\cdot\! \vec{I}=a(I_zS_z + I_xS_x + I_yS_y). 
\eeq 
When the non-secular components are taken
into account, the eigenvalues of the resulting Hamiltonian reveal a
further splitting of $a^2/\w_e = 26$~kHz~$= 0.9$~$\mu$T. Each hyperfine line (marked with $M_I=0$ and $\pm 1$ in
\Fig{cwEPR}(B)) involves the three
allowed electron spin transitions $\Delta M_S=1$ within the $S=3/2$
multiplet. These electron spin transitions remain degenerate for
$M_I=0$, as seen in \Fig{cwEPR}(D), but split into three lines
(with relative intensities 3:4:3) for $M_I=\pm 1$, as seen in
Figures~\ref{cwEPR}(C) and (E). The observation of this additional
splitting is only possible because of the extremely narrow EPR
linewidth $< 0.3~\mu$T. This linewidth is still limited by the
resolution of the spectrometer, in particular, magnet stability and
homogeneity. Similar second-order
splittings have been reported for the related spin system in the
endohedral fullerene $^{31}$\patc, which has $S=3/2$ coupled with
$I=1/2$ \cite{knapp98}. The hyperfine splitting in \patc~is
substantially larger, and hence so is the second order correction.
This second order splitting also leads to a profound modulation of the electron spin echo, with a principal component at 26~kHz for \natc~\cite{eseem05}. 

This second order splitting results in a non-uniform spacing of the levels, and therefore it should be possible to
selectively address the three individual electron spin transitions
(at least, for the outer hyperfine lines). However, at X-band microwave frequency, the splitting of
0.9~$\mu$T is far too small to be of practical use. The
homogeneity of the magnet in our pulsed EPR machine is not good
enough to resolve the splitting, and a pulse would have to be
1000 times longer than usual ($> 50~\mu$s) in order to be suitably
selective. Note, however, that the splitting scales reciprocally with microwave frequency and therefore a greater splitting is expected if we operate at
lower microwave frequencies.

The direct observation of a $^{13}$C-hyperfine interaction of 1.3~$\mu$T~=~36~kHz is shown in
\Fig{cwEPR}(D). The different peaks correspond to cages
with different numbers of  $^{13}$C atoms. Given the natural
abundance of  $^{13}$C (1.07$\%$), we can calculate the expected
peak intensities for the cases of zero, one or two  $^{13}$C atoms
on a \csixty~cage: (57$\%$, 30$\%$ and 12$\%$, respectively), which
are in good agreement with the observed spectrum. The measured
hyperfine splitting is consistent with the value obtained from a
$^{13}$C ENDOR study~\cite{weiden99}. The strength of the hyperfine interaction gives an indication of the
electron spin density on the nucleus. Measuring the hyperfine
coupling therefore provides us with an estimate of spin delocalization of the nitrogen atom over the C60 cage. This
isotropic hyperfine coupling constant of 34~kHz corresponds to an
approximate 2$\%$ transfer of spin density from the nitrogen atom to
the C$_{60}$ molecule\footnote{A unit spin density on the \csixty~cage is expected to produce a \cthirteen~hyperfine interaction of approximately 1.5~MHz~\cite{knapp97}}. 

In other candidate endohedral species such as Sc@C$_{82}$, the anisotropy of the hyperfine interaction and complex nature of the bond between the bound atom and the cage make the interpretation of EPR spectra more challenging. Using density functional theory we have modeled the electron charge and spin distributions in Sc@C$_{82}$, where the bond with the cage is partly covalent and partly ionic and most of the electron spin density is distributed around the carbon cage. The
anisotropy is attributed to 5\% occupation of the Sc d$_{yz}$ orbital~\cite{morleysc}.

\begin{figure}\centerline
{\includegraphics[width=3.5in]{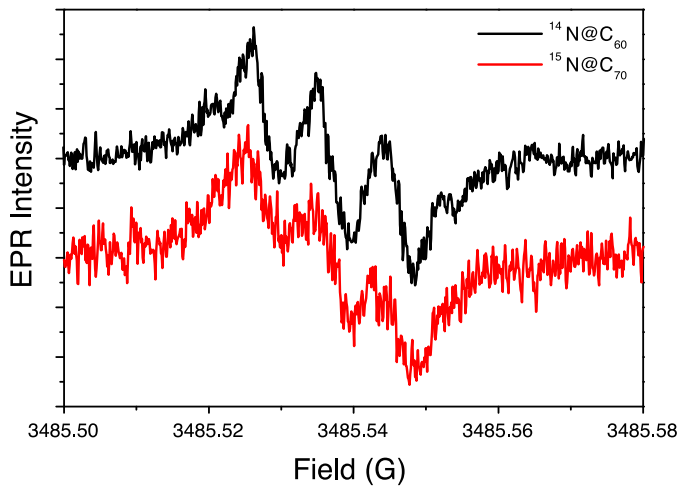}}
\caption{Zoom-in of the CW EPR spectrum for the low-field line in $^{14}$\natc~and $^{15}$\natcseventy. By chance, different factors conspire to give an identical second-order hyperfine splitting.
 } \label{nc70split}
\end{figure}

\subsection{$^{15}$\natcseventy: Effects of isotope and cage}

Different isotopes of nitrogen can be used during the implantation process (indeed, trace amounts of $^{15}$\natc~were detected when the molecule was first synthesised~\cite{Murphy1996}), and a nitrogen atom can be equally well encapsulated within a slightly larger, ellipsoidal cage: \cseventy. The isotopic change has a dramatic effect on the hyperfine coupling constant, whilst the change of cage size provides a more subtle shift. The nuclear gyromagnetic ratio of \nfifteen~is 1.4 times larger than \nfourteen, indicating a hyperfine coupling constant of $5.66 \times 1.4 = 7.92$~G. However, the larger cage of \cseventy~relaxes the compression of the electron cloud, reducing the hyperfine constant by appoximately 5$\%$~\cite{dietel99}.  As \nfifteen~has $I=1/2$, the CW EPR spectrum shows two principal lines, with second-order hyperfine splittings observable on both hyperfine lines ($\delta$ in this case is $a^2/(2B)$). By coincidence, the resulting second-order splitting matches that for \natc~almost exactly, as revealed in \Fig{nc70split}.

\subsection{Quantum coherence in \natc}

\begin{figure}[t]
\centerline {\includegraphics[width=6in]{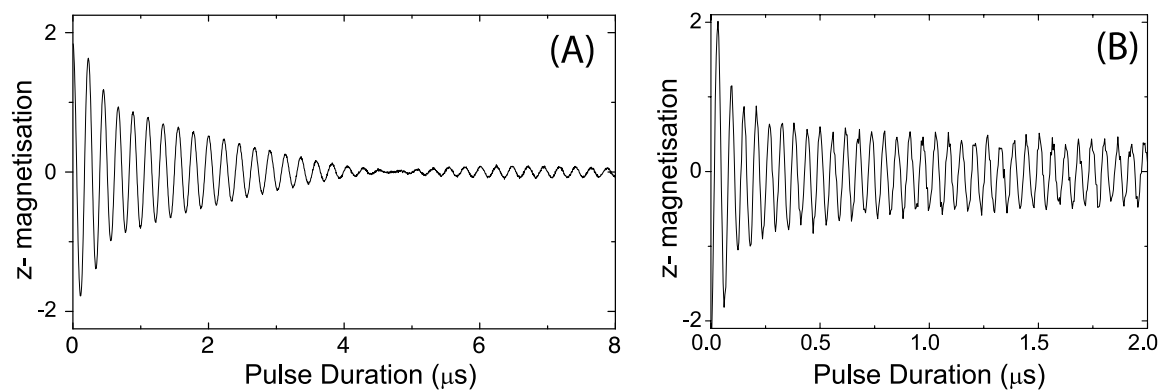}}
\caption{Electron spin Rabi oscillations for \natc~in \cstwo~at 190
K. Sample dimensions: (A) 15~mm long, 4~mm diameter; (B) 7~mm long, 1~mm diameter. The observed decay is due to inhomogeneity of the applied microwave field, and can thus be suppressed by reducing the physical dimensions of the sample~\cite{morton04} .} \label{LongPulse}
\end{figure}

The ability of \natc~to faithfully store quantum information is characterised by the relaxation time $T_1$ and the coherence time $T_2$. These have been studied in a range of different environments, and yielded measurements of $T_2$ up to 0.25~ms in liquid \cstwo~solution at 160K~\cite{relaxcs2}. With nutation period 32~ns in a typical X-band EPR spectrometer, this T$_2$  time corresponds to more than $10^4$ electron spin Rabi oscillations, examples of which are shown in \Fig{LongPulse}.

In addition to choosing systems with long $T_1$ and $T_2$ times, it is also essential to evaluate and minimise the errors associated with qubit logic gates when judging quantum computing implementations. We have shown that even in an EPR system with a $10\%$ systematic error in single qubit operations, composite pulses (such as BB1) enable fidelities between 0.999-0.9999 to be achieved, and the theoretical limit of this method should be even better~\cite{mortonbb1}. Along with the measured coherence times, this meets commonly accepted requirements for fault-tolerant quantum computation~\cite{steane03}.

\subsection{The nuclear spin as a resource}
\begin{figure} \centerline
{\includegraphics[width=4in]{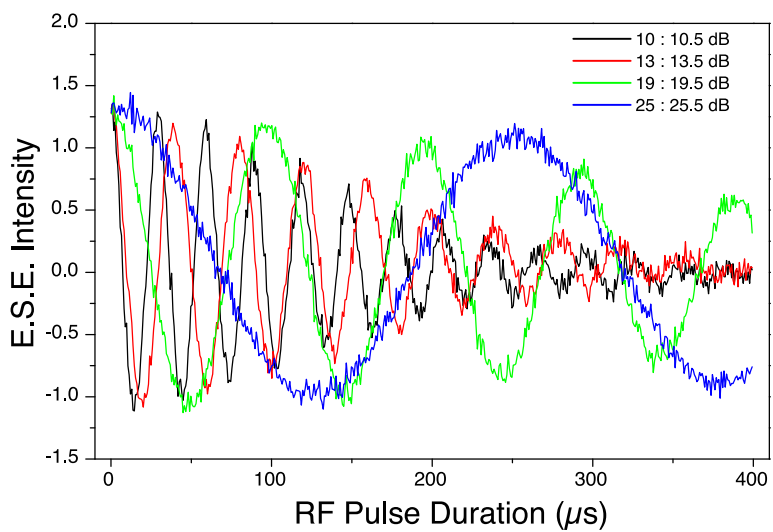}} \caption{The nuclear Rabi frequency can be controlled by changing the power of the RF driving field. To maximise detection efficiency, the two RF transitions at the $M_S=\pm3/2$ manifolds are driven simultaneously (at 22.597 and 24.781~MHz)}
\label{rabipowers}
\end{figure}

In addition to providing the hyperfine coupling that will enable the distinction of different types of qubit in a computer containing $^{15}$\natc~and $^{14}$\natc~fullerene subunits, the nuclear spin is a resource in its own right. Capable of even longer storage times of quantum information than the electron spin, its state could be swapped with the electron spin when a computation is not taking place. The nuclear spin can be manipulated by RF pulses, as shown in the nuclear Rabi oscillations in \Fig{rabipowers}, and the presence of the electron spin can be exploited to generate ultra-fast phase gates, and to further protect the nuclear spin from unwanted interactions by \emph{bang-bang} decoupling~\cite{bangbang, viola98}.

\section{Optical mechanisms for single spin measurement and manipulation}

A vital requirement of the proposed endohedral electron
spin approach is to be able to measure single spins. All the ESR measurements so far of spins in fullerenes have been performed using the free induction decay of ensembles containing of order $10^{14}$ molecules. There are several candidate technologies for measuring single electron spins. Direct measurements of small magnetic fields can be made by micro-SQUIDs (superconducting quantum interference devices), but current sensitivity is limited to $\Delta m_S=30$, which corresponds to the flipping of 30 electron spins~\cite{pakes01}. Magnetic resonance force microscopy (MRFM) has been proposed for a nuclear-spin based quantum computer~\cite{berman00}, and has been used to detect an individual electron spin in silicon dioxide, with a spatial resolution of 25~nm~\cite{rugar04}. This beautiful experiment allows a direct detection of a single spin, but the measurement process is slow. STM assisted EPR allows modulation of the tunnelling current at the Larmor precession frequency of single spins to be detected, but this is not a vector measurement, and therefore cannot make a projective measurement of an electron spin qubit~\cite{durkan02}. Single-shot measurements of single electrons in GaAs quantum dot (lithographically defined within a 2D-EG) are possible using a quantum point contact ~\cite{elzerman04}, thus providing proof-of-principle that electrical measurements of single spins in solid state devices are possible. A single fullerene molecule can be coupled to a pair of electrical contacts, as demonstrated in molecular transport experiments~\cite{parktransport}. Given sufficiently low temperatures and magnetic fields, it may be possible to use Coulomb blockade in such a molecular-nanoelectronics hybrid device to measure an endohedral electron spin~\cite{elste04}.

\begin{figure}\centerline
{\includegraphics[width=4.2in]{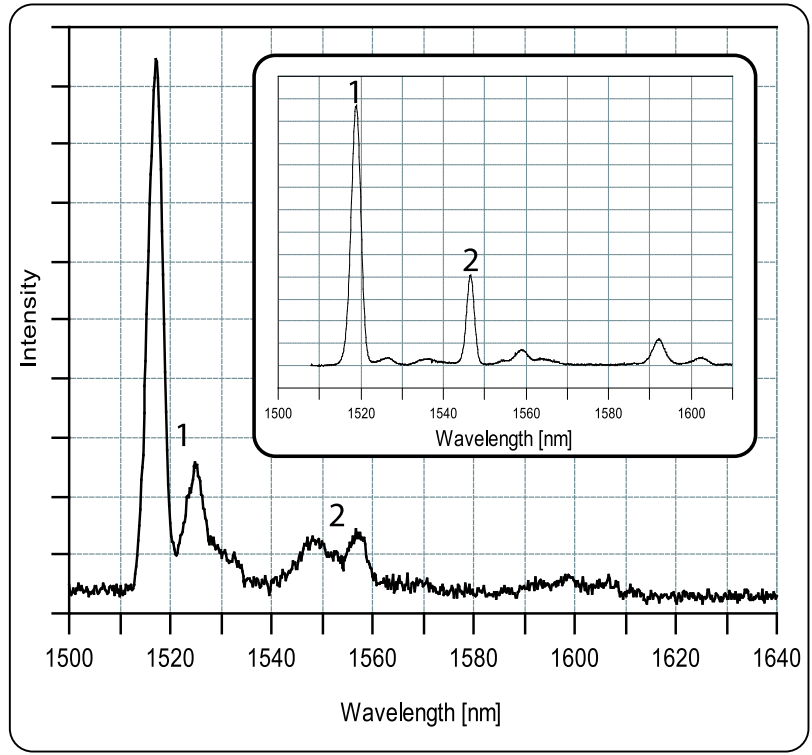}}
\caption{$\mathrm{1.5\,\mu m}$ photoluminescence observed from
$\mathrm{Er_3N@C_{80}}$ at $\mathrm{4.2\,K}$ at  $\mathrm{19.5\,T}$. Inset spectrum shows $\mathrm{0\,T}$ data. Feature 1 and 2 both split linearly with field. Splitting of peaks with field indicates that the magnetic degeneracy of this system can be lifted, indicating the potential for ESR activity in this class of materials.
 } \label{photoLum}
\end{figure}

An alternative to electrical measurement of a single spin would be optical measurement. In \emph{Optically Detected Magnetic Resonance} (ODMR) the luminescence of a sample is observed under a magnetic field and used to detect electron spin.  In titanium-doped silicon carbide, radiative transitions between two electronic spin states (S=1 and S=0) on the titanium atom have been optically detected~\cite{lee85}. It may be possible to couple the endohedral spin to a solid-state magnetic optical dipole such as nitrogen-vacancy centres in nanocrystalline diamond~\cite{charnock01,twamley03}, which have themselves formed the basis of a QIP proposal~\cite{lukin00} and in which a single spin has been optically measured~\cite{wrachtrup04}. Spin to photon conversion has also been demonstrated in InAs self-assembled quantum dots~\cite{kroutvar04}. Direct optical detection of spin in N@C$_{60}$ and P@C$_{60}$ appears elusive, because the transition to the first excited state in atomic nitrogen falls far into the ultraviolet, which is beyond the capabilities of standard optics apparatus and well inside the primary absorption of the C$_{60}$ cage.

The lanthanide metallofullerenes, in particular the erbium-doped fullerenes, offer greater promise for optical readout. The singly-doped variant Er@C$_{82}$ displays paramagnetic resonance \cite{San01}, but does not exhibit detectable luminescence \cite{Mac97}, as in this case unfilled cage molecular orbitals lead to strong absorption at the 1.5$\,\mu$m Er$^{3+}$ first-excited-state $^4I_{13/2}$ to ground-state $^4I_{15/2}$ transition wavelength. The doubly-doped Er$_2$@C$_{82}$ exhibits 1.5$\,\mu$m photoluminescent emission characteristic of Er$^{3+}$ $4f$-electron transitions \cite{Hof98, Mac97}, but the pairing of the endohedral spins leads to a spin-silent molecule \cite{San01}. At cryogenic temperatures, the Er$_2$@C$_{82}$ spectrum appears as a set of 8 well-resolved lines owing to lifting of the $m_J$ degeneracies inherent in the $^4I_{13/2}$ and $^4I_{15/2}$ manifolds by the local crystal field. This photoluminescence process is interpreted as absorption of the exciting visible laser source into the strongly absorbing cage states followed by nonradiative relaxation from these states to the ion, and then \emph{intra} the ion, between the ions or via the cage to the $^4I_{\frac{13}{2}}$ manifold lowest level, and then luminescent decay. At elevated ($\mathrm{300\,K}$) temperature, phonon broadening and thermal population of the upper levels prevent observation of the characteristic lines. The crystal-field splittings of around $15\,\mathrm{cm^{-1}}$ are large compared to crystal hosts, but the close proximity of the erbium ions and the charged cage make the strength of the interaction credible. The lifetime of the lowest $^4I_{\frac{13}{2}}$ level is less than $5\,\mathrm{\mu s}$, giving a quantum efficiency of less than 1\%.

Another class of erbium-doped fullerenes, the so-called TNT fullerenes, which are composed of a planar trigonal tri-lanthanide nitride group enclosed in a 78, 80 or 82-carbon cage, offer additional candidates, $\mathrm{ErSc_2N@C_{80}}$, $\mathrm{Er_2ScN@C_{80}}$ and $\mathrm{Er_3N@C_{80}}$~\cite{Ste99}. Similar cage-mediated photoluminescence measurements at $4.2\,\mathrm{K}$ of these erbium-scandium cluster fullerenes produces an 8-line Er$^{3+}$ $\mathrm{1.5\,\mu m}$ spectrum~\cite{Mac01}. At $77\,\mathrm{K}$, the lines are again broadened due to phonon effects, and thermal population of the upper $\mathrm{^4I_\frac{13}{2}}$ manifold leads to the appearance of further lines (hot bands). The fluorescence lifetime is $2\,\mathrm{\mu s}$ giving a a quantum efficiency of $10^{-4}$. All these photoluminescence measurements have been performed by exciting the ionic states via the cage states; because of the uncontrolled and complex relaxation pathways involved this is a highly unsuitable process for performing the delicate and precise coherent ionic manipulations required of a candidate readout scheme. However, it is also possible to manipulate the ionic states directly, using the same absorption-free wavelength region. We have recently performed direct optical excitation of individual ground-state to first excited manifold ionic transitions in the erbium TNTs~\cite{jon05a}. This unlocks not only the potential to survey the states of the upper manifold with a view to identifying useful readout transitions, but also to excite these transitions directly, coherently, and selectively.

The Er$^{3+}$ ion, being a Kramers ion, maintains a twofold degeneracy in its quantum states, even under complete crystal-field splitting. In the presence of a large magnetic field, this degeneracy may be lifted and the observed transitions split. The application of magnetic field in a sufficiently crystal-field-split case can produce an effective spin-$1/2$ system, a qubit candidate. We have applied a $19.5\,\mathrm{T}$ magnetic field during a luminescence measurement on this TNT system (see \Fig{photoLum})~\cite{jon05b}. The spectrum is observed to split indicating that the ionic states are responsive to external magnetic fields, and confirming that the ground state is indeed a Kramers doublet. By encoding quantum information in this pseudo-spin-$1/2$ qubit, and then directly exciting spin-selective luminescent transitions, it may become possible to optically detect a spin state in this or similar species of endohedral fullerene.

\section{Toward one-dimensional arrays for QIP}

\subsection{Fullerene Dimers}

The considerable wealth of experimental information that has been gained from one qubit system based upon N@C$_{60}$ gives encouragement to prepare multiple qubit systems.  An immediate extension of ensemble measurements on single fullerene systems is fullerene dimers, i.e. a bonded pair of fullerenes. If both cages contain spin-active species then this would give rise to a two-qubit system.  

The directly bonded fullerene dimer, C$_{120}$, can be readily synthesised through a high-speed vibration milling (HSVM) technique.  Despite the vigorous nature of this synthesis, the dimer N@C$_{60}$-C$_{60}$, in which one fullerene cage contains a nitrogen atom \cite{goedde01}, confirms the resilience of the encapsulated nitrogen to chemical bond formation on the surface of fullerene cage.  Asymmetric 
C$_{60}$-C$_{70}$ dimers can be prepared and isolated, offering the possibility of directly bonded A-B spin active dimers. 

\begin{figure}\centerline
{\includegraphics[width=4.5in]{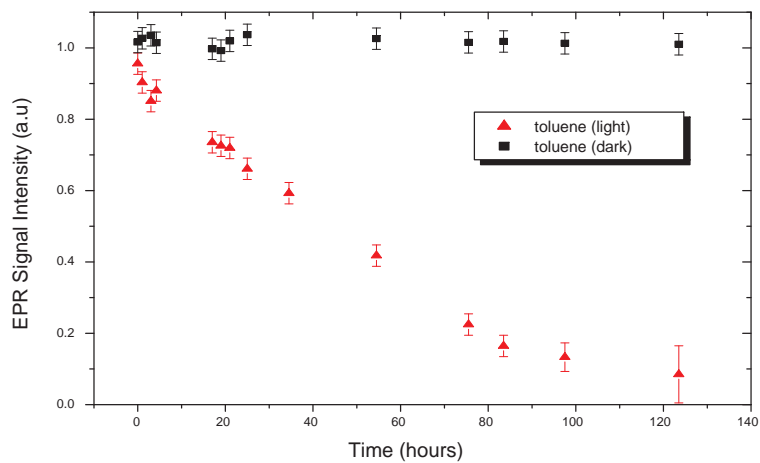}}
\caption{A loss of nitrogen spin signal in the ESR spectrum of toluene solutions of N@C$_{60}$O under vacuum is observed over several days upon exposure to ambient light.  In the absence of light N@C$_{60}$O is stable and no loss of spins is observed.} 
\label{KP_nc60Decay}
\end{figure}

The synthesis of the oxygen-bridged dimer, C$_{120}$-O, by reaction of the fullerene epoxide C$_{60}$O with C$_{60}$ \cite{Murray,Lebedkin, Tokuyama} provides, if the unfunctionalised C$_{60}$ is present in a large excess, a route to selective dimerisation.  We have recently prepared the epoxide functionalised endohedral fullerene N@C$_{60}$O~\cite{bigPreprint}, which holds potential for dimerisation with $^{15}$N@C$_{60}$ to yield a $^{14}$N-$^{15}$N two-qubit system.  The lower thermal and photolytic stability of N@C$_{60}$O synthon versus underivatised N@C$_{60}$ are thought to stem from the highly reactive epoxide ring (Figure \ref{KP_nc60Decay}).

\begin{figure}\centerline
{\includegraphics[width=5.in]{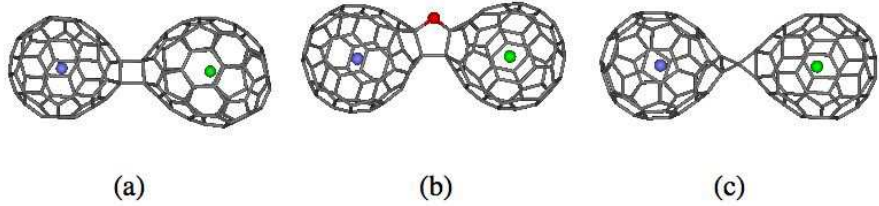}}
\caption{Fullerene dimers. Schematic representation of the molecular structures of (a) $^{14}$N@C$_{60}$-$^{15}$N@C$_{70}$, (b) $^{14}$N@C$_{60}$-O-$^{15}$N@C$_{60}$ and (c) $^{14}$N@C$_{60}$-C-$^{15}$N@C$_{60}$. } 
\label{3dimers}
\end{figure}

 In a related synthetic strategy, the endohedral fullerene derivative $^{14}$N@C$_{61}$Br$_2$, which has been recently prepared in our laboratories \cite{Sambrook}, could be reacted with an excess of $^{15}$N@C$_{60}$ to yield a two qubit carbon bridged C$_{121}$ dimer.  This dimerisation route offers enhanced stability of both the N@C$_{61}$Br$_2$ intermediate and the dimer product in comparison to N@C$_{60}$O and the furan bridged C$_{120}$O dimer, respectively. Schematic representations of the molecular structures of these endohedral fullerene dimer systems are shown in Figure \ref{3dimers}.  These two stage synthetic routes for dimer production offer (i) the short synthetic route requiring functionalisation of only one carbon cage and (ii) the close proximity of both encapsulated nitrogen atoms in the final product, thus maximising interaction strength and ultimately the rate of entanglement.

Various fullerenes are being developed for photovoltaics, optical devices, including the rapidly developing OLED field, nanoelectronics and artificial devuices that mimic photosynthesis \cite{Herranz, Roncali}.  Many mono- and bis-fullerene systems have been designed towards these goals, a common feature being the presence of highly conjugated, optically active species such as porphyrins, $\pi$-conjugated oligomers, metal coordination compounds and tetrathiafulvalene (TTF). All of these systems potentially hold value for QIP purposes; highly conjugated linkers between cages may provide a route for an exchange interaction between two endohedral fullerene qubits whereas optically active species may allow control of this qubit entanglement. Non-covalent interactions may be useful for the assembly of large arrays of endohedral fullerenes for QIP.  Such interactions could include hydrogen-bonding, $\pi$-$\pi$ stacking interactions, coordination chemistry and solvophobic effects.  Although weak in comparison to covalent bonds, it is well-established that highly stable assemblies can be achieved through the cooperative effect of multiple interactions. The risk of incomplete or incorrect arrays is reduced by the inherent Ôerror-correctingÕ ability of these thermodynamically driven assembly processes ~\cite{Lindsey}.  

\subsection{``Peapod'' Nanotubes}

To make a quantum circuit larger than the two qubits afforded by a fullerene dimer, fullerene structures need to be scaled to larger arrays. Fullerenes can be assembled into ordered arrays in single-walled carbon nanotube (SWNT). Self-assembled molecular networks have advantages over individually placing an atom in a trap or implanted in a surface, as the network periodicity and geometry are dictated by well-defined molecular interactions, usually noncovalent bonding \cite{Theobald}.  Fullerenes spontaneously enter open nanotubes upon heating to form ÔpeapodsÕ \cite{monthioux}. Nanotubes provide unique systems suitable for electronics, showing one-dimensional ballistic electron transport \cite{mann,javey,todorov} and remarkably long spin coherence lengths \cite{groeneveld}. Nanotubes can be side-gated with self-aligned, regularly spaced metal electrodes tens of nanometers wide \cite{Chhowalla}. Fullerenes in a SWNT locally alter the electronic states of the SWNT \cite{Shinohara2005nat,horsfield}. If the electronic states of the nanotube and fullerene are coupled, then this offers a mechanism for qubits to interact over distance through the nanotube. SWNTs have been grown up to millimeters in length \cite{mizuno,bernardi}, which could hold several thousand qubits arranged into a one-dimensional, self-assembled chain. 

\begin{figure}\centerline
{\includegraphics[width=4.in]{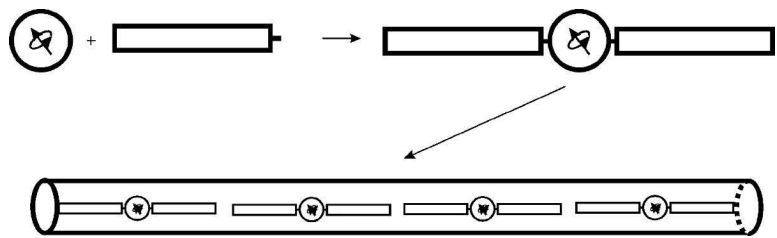}}
\caption{Scheme for forming periodic arrays of spaced spin-active fullerenes in a SWNT. A spin-active fullerene is bis-functionalised and then inserted in a SWNT using a low temperature filling method to avoid degradation of the electron spin and the functional groups.} 
\label{DB_tubeSpacing}
\end{figure}

\begin{figure}\centerline
{\includegraphics[width=4.in]{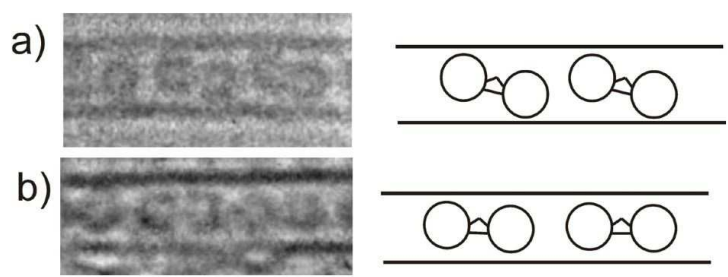}}
\caption{HRTEM micro-graphs of C$_{120}$O in SWNTs. (a) shows that dimers tilt in wider 
nanotubes, whereas in (b) narrower nanotubes, they are in linear arrays.} 
\label{DB_tubeDimer}
\end{figure}

For fullerene spins to be controlled locally by gates, they need to be spaced at distances suitable for each gate to address a single fullerene. We have studied the functionalization of N@C$_{60}$, described above, and found that it is possible to attach functional groups to N@C$_{60}$ without loss of the nitrogen spin. These groups can be further altered to create spacers so that fullerenes can have a controllable distance between them. Bis-functionalised fullerenes would assemble with regular periodicity inside SWNTs (Figure \ref{DB_tubeSpacing}). An alternative method of addressing spins and transferring quantum information is to use a Ôglobal addressingÕ scheme originally proposed by Lloyd \cite{lloyd93} and later refined by Benjamin \cite{benjamin}, as we presently discuss. The beauty of marrying global addressing with self assembly is that arbitrarily large quantum computers would be possible with minimal effort on the part of the architect, as long as the basic interactions between spins are characterized. For example, such self-assembly could be achieved by encapsulating fullerene dimers that have a preferential orientation for entry (creating an ABAB... array). We have encapsulated C$_{120}$O fullerene dimers in SWNT (\Fig{DB_tubeDimer}): the dimers lie flat in narrow SWNTs, and tilt to maximise the van der Waals interaction in wider SWNTs.

\begin{figure}\centerline
{\includegraphics[width=4.in]{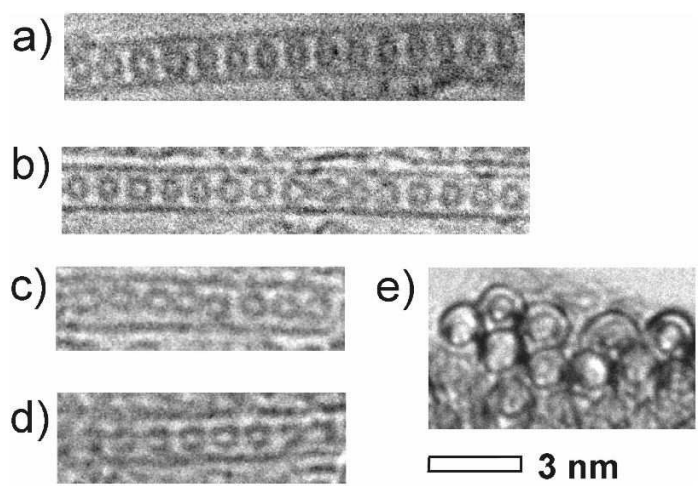}}
\caption{HRTEM micrographs of C$_{70}$@SWNT in (a) and (b) standing orientation and (c) and (d) in the lying orientation.  (e) The end of a bundle of C$_{70}$@SWNTs, showing complete filling to the end of the nanotubes.} 
\label{DB_tubeVariety}
\end{figure}

Fullerenes inside nanotubes are quasi-one dimensional systems; the fullerenes can orient themselves relative to the nanotube sidewall and to their nearest neighbours. We have shown that rugby ball-shaped C$_{70}$ will lie along the axis of 1.36 nm SWNTs and will stand in 1.49 nm SWNTs (see \Fig{DB_tubeVariety}) \cite{c70orient}. For spherical C$_{60}$, increasing the nanotube diameter causes fullerene arrays to go from a linear chain to a zig-zag, followed by a double helix (see \Fig{nc60cartoon2}), then a two-molecule layer \cite{c60phases}. This observed transition matches theoretical predictions very well \cite{girifalco}. In both cases, ordering of these arrays is strictly controlled by non-directional van der Waals interactions. Electrostatic interactions give us a further handle to orient metallofullerenes such as Ce@C$_{82}$ inside SWNTs \cite{andrei04}. Other directional interactions, such as hydrogen or covalent bonding, can encourage formation of ordered arrays of fullerenes. 

We have developed a method for inserting molecules into SWNTs at low temperature and in an inert environment to preserve molecular functionality using supercritical fluids. This technology is applicable to a wide variety of fullerenes, including fullerenes with spacer groups and groups that hydrogen bonds \cite{PoliakoffJMC,britzCC2005,britzCC2004}. Extending this approach, we can make novel covalently bonded structures by inserting highly reactive fullerene epoxide, C$_{60}$O, into SWNTs using supercritical carbon dioxide and then heating the resulting peapods to form a (C$_{60}$O)$_n$ polymer. The one-dimensional (C$_{60}$O)$_n$ polymer inside the SWNT has a similar structure to C$_{60}$ peapods, but each fullerene is covalently bonded to two nearest neighbours. (C$_{60}$O)$_n$ in the bulk forms a three dimensional, branched, disordered polymer, demonstrating that the inside of a SWNT is a more controlled environment to design arrays of fullerenes. 

This low-temperature filling technique can be used for forming spin active arrays, since functional groups remain intact after insertion of functionalized fullerenes into SWNTs. We have inserted 1\% N@C$_{60}$/C$_{60}$ into SWNTs in high yield by cycling solvent pressure, to attain peapods in 70\% yield, as confirmed by HRTEM imaging. The nitrogen spin is preserved after encapsulation, as shown by EPR of the peapods suspended in CCl$_4$ \cite{morleyThesis}. The (N@C$_{60}$/C$_{60}$)@SWNT has an EPR linewidth broader than would be expected for a 1D chain of N@C$_{60}$/C$_{60}$ interacting solely by dipolar spin coupling. This observation could be due to slightly different environments in different nanotubes causing an inhomogeneous broadening or another coupling mechanism of the spins, such as via the nanotube wall. 

\subsection{Static and flying qubits in nanotubes}

As well as providing structural support for arranging fullerenes in one-dimensional arrays, carbon nanotubes may provide new ways of controlling the interactions between fullerenes. Electrons may be confined in a semiconducting nanotube using electrostatic gates to create one or more potential wells~\cite{mason04}.
If such a well is sufficiently shallow and narrow it will bind only
a single electron~\cite{entangler}. Nanotubes have
very high circumferential confinement, which, so long as the confining potential along the tube is sufficiently deep, leads to large excitation energies for the confined electronic charges. This inhibits charge fluctuations due to disorder and offers well-defined spin-qubits (`static' qubits) which can be manipulated using electric fields, magnetic fields and their mutual Coulomb interaction. Single conduction electrons (`flying' qubits) can be added one at a time to semiconducting single-wall nanotubes, using
a turnstile device~\cite{KouwenhovenPRL}. They may also be selected for their spin orientation, using a magnetic contact, a quantum-dot filter \cite{recher} or a Zener filter \cite{zener}. Such spin-polarized electrons can interact with the confined
electron spins, exchanging quantum information and inducing an effective interaction between pairs of static qubits. The spin of the transmitted qubits may be analysed by a combination of spin filter and single electron detector, allowing characterisation of the type of interaction induced.

Instead of using gates to confine electrons to provide the static qubits, peapod structures offer electron spins that are already localised in the molecular orbitals. Interactions
between adjacent qubit bearing electrons can be understood in
terms of virtual charge fluctuations which give rise to weak antiferromagnetic
Heisenberg coupling between spins in semiconducting nanotubes. For
a double-well system this gives a fully entangled singlet ground state.
In metallic or doped semiconducting nanotubes there is a competition between the tendency to form Kondo singlet ground states between the static qubit and the Fermi sea of electrons in the nanotube~\cite{hewson93}, and an enhanced Heisenberg/RKKY type 
interaction~\cite{ruderman54, kasuya56, yosida57}, which may be modulated
in both magnitude and sign by changing the Fermi energy with gates. The versatility of the fullerene and nanotube materials offers scope for finding a regime in which RKKY will dominate over Kondo, with corresponding potential for fast controlled two-qubit gates.

\begin{figure}
\centerline
{\includegraphics[width=5.in]{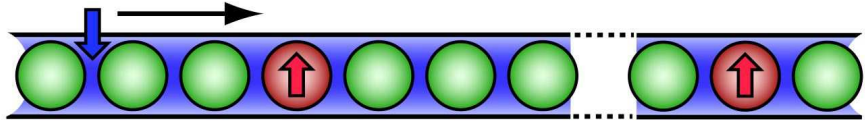}}
\caption{Schematic of proposed peapod system in which a single fullerene with
an excess electron (red, static spin-qubit) becomes entangled with an injected
electron of opposite spin (blue, flying spin-qubit). Empty spacer fullerenes surround the spin-active fullerene; further down the tube there may be other spin-active fullerenes, or alternatively a spin filter to enable spin-to-charge conversion for measurement.} 
\label{JJ_spinsInTube}
\end{figure}

In semiconducting single-wall carbon nanotubes or peapods, we have the further
possibility of using the correlations between the spin
of a single propagating electron and that of a bound
electron as a resource for quantum information processing. These are induced by a combination of Coulomb
repulsion and Pauli exclusion~\cite{wires, entangler}.
For total $S_{z}=0$, there is an effective antiferromagnetic exchange
interaction between the spins of the incident and bound electrons
which induces entanglement between them. Detailed calculation for
the case when an electron is bound by an electrostatic gate shows
that the entanglement of the asymptotic state after scattering may be tuned by choice of initial kinetic
energy of the incident electron. This can induce maximal entanglement
at two specific energies, corresponding to the singlet and triplet resonance
energies. The high confinement around the tube places restrictions
on the confining well in order that the asymptotic states after interaction
leave a single bound electron in the well. For relatively wide wells,
such as those produced by metallic gates deposited using electron-beam
lithography, the well has to be shallow to avoid ionisation of the
bound electron into the conduction band. In this regime the two electrons
are strongly correlated when they are both in the well. This gives rise to a weak antiferromagnetic
Heisenberg coupling between the spins whose strength is much less
than the resonance widths -- and only partial entanglement is possible.
To increase the effective spin-spin interaction, the well must be
made narrower and deeper. Even in this highly confined regime, electron correlations are still important
due to the small effective Bohr radius, itself a consequence of the
small dielectric constant. These effects may be achieved in a dilute peapod, containing relatively isolated spin bearing fullerenes separated by empty fullerenes (\Fig{JJ_spinsInTube}), or by spacer molecules (cf. \Fig{DB_tubeSpacing}), to create a structure suitable for exploiting these ideas for static-flying qubit entanglement. 

The peapod structures are so rich in properties that there may be other factors to be understood and controlled, such as orbital degeneracy with associated Jahn-Teller effects on the cage, the relative positions
of the nanotube conduction band and the fullerene HOMO orbital, and the
magnitude of the coupling between nanotube and fullerene. The latter
may be estimated from DFT calculations that show hybridisation energies of up to a few tens of meV~\cite{hybrid},
and Coulomb repulsion energies of doubly-charged fullerenes of  order $2-3$~eV. Each of these offers the potential for further ways of manipulating the entanglement of the electron spins.


\section{Scaling up to large interacting systems: 2D-Arrays on surfaces}

\begin{figure}\centerline
{\includegraphics[width=4in]{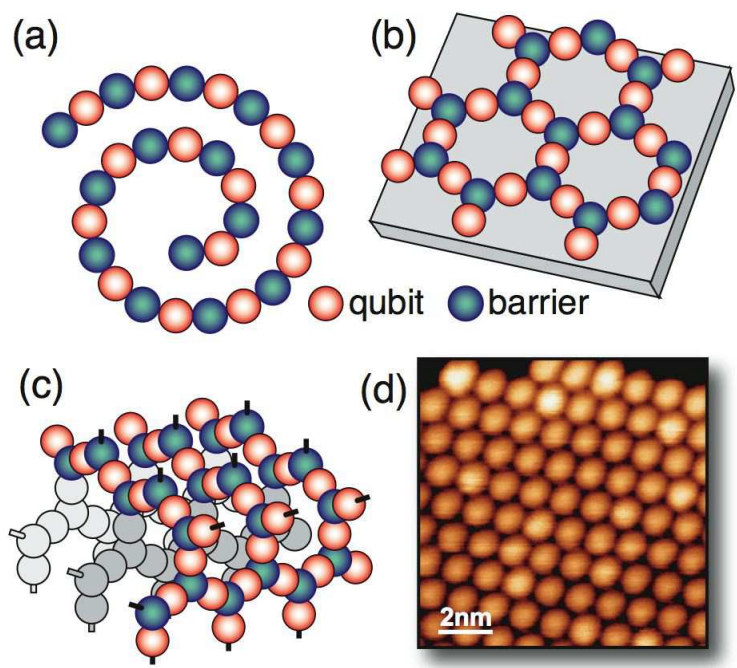}}
\caption{Arrays of different dimensionality. (a)-(c) Images showing various periodic array architectures which can support QIP \cite{benjaminNJP}. The spheres represent two different types of spin qubits (for example, two species of endohedral fullerene); one species acts as a barrier while the other represents the logical qubits \cite{benjaminNJP,benjaminAndBose,benjamin05}. (d) STM image taken at liquid nitrogen temperature of an island of Er$_3$N@C$_{80}$ molecules at -2.0~V on a Ag/Si(111) surface.} 
\label{KP_Leigh}
\end{figure}

Ultimately for a mature QIP technology, it will be desirable to create two-dimensional (or higher) structures. Although the power of a quantum computer is such that, in principle, even a one-dimensional device can profoundly outperform any classical machine on suitable tasks, nevertheless issues such as fault tolerance motivate work towards structures with better connectivity. Molecular structures can of course produce two- and three-dimensional arrays with long range order. There are two related concerns that arise when when considers such structures as QIP architectures.  The first is that of {\em addressability}: given the nanometre scale of the array period, can one hope to manipulate individual qubits and their interactions? In few-qubits systems this is not an issue since each element may have a unique signature frequency, but in large arrays there will be many identical elements. Fortunately there is an available solution in the form of {\em global control}: one can show \cite{lloyd93,benjamin, benjaminAndBose, BrendonPreP} that pulses sent to an entire array can have a net effect in just one place. Secondly, one may be concerned that interactions between nearby elements cannot be {\em switched} ``on'' and ``off'', regardless of whether such switching is local or global. Here also there are solutions; one idea we have examined involves using some subset of the physical qubits as barriers, rather than information bearing units \cite{benjaminNJP, benjaminAndBose,benjamin05, BrendonPreP}. As shown in \Fig{KP_Leigh}, this idea is compatible with one, two or three-dimensional arrays.

There are several physical mechanisms by which one can form arrays of fullerenes to realize such architectures. Two dimensional supramolecular assemblies can be made on surfaces by exploiting non-covalent interaction between the constituent molecules. Such molecular networks form porous structures that can act as hosts for fullerene molecules. The number and arrangement of the guest fullerene molecules is largely controlled by the size and shape of the pores of the host network. Fullerene heptamers with hexagonal packing have been formed in a perylene tetra-carboxylic di-imide (PTCDI)-melamine network on a silver-terminated silicon surface \cite{Theobald}. Single C$_{60}$ molecules have been incorporated in a trimesic acid (TMA) molecular network on graphite \cite{Griessl}. Metal-organic coordination networks have also been used as hosts for guest C$_{60}$ molecules on a Cu surface \cite{Stepanow}. Even without the presence of a templating network, fullerenes deposited on a substrate tend to arrange themselves into domains with hexagonal packing \cite{Leigh}. Figure \ref{KP_Leigh} shows a filled-states STM image taken at liquid nitrogen temperature of an island of Er$_3$N@C$_{80}$ molecules at -2.0~V on a Ag/Si(111) surface. These examples demonstrate how it is possible to arrange endohedral fullerenes in an ordered 2-D pattern, offering new possibilities for computer architectures, and fault tollerance around defective regions through the availability of alternative pathways.

\section{Conclusions}
In this paper, we have reported on a wide range of research, both theoretical and experimental, which we have undertaken in order to exploit fullerenes as a component for a quantum information technology. We have described our successes in establishing the suitability of a family of endohedral fullerenes as a unit for storing and manipulating quantum information, and we have discussed our research into optical manipulation of these molecules. We then reported on various lines of research we have undertaken to synthesize arrays: both small dimer structures and extended scalable arrays e.g. inside nanotubes. In this context we have discussed our theoretical work on interactions and on array architectures. Throughout we have sought to highlight the questions that remain to be answered. Although there are many such questions, we believe that our work has already established that these beautiful molecules are indeed highly promising candidates for future quantum technologies. 

\section{Acknowledgements} 
We thank the Oxford-Princeton Link fund for support. This research is part of the QIP IRC www.qipirc.org (GR/S82176/01). GADB thanks EPSRC for a Professorial Research Fellowship (GR/S15808/01). AA and SCB are supported by the Royal Society. JHJ acknowledges support from the UK MoD. JJLM is supported by St. John's College, Oxford. Work at Princeton was supported by the NSF International Office through the Princeton MRSEC Grant No. DMR-0213706 and by the ARO and ARDA under Contract No. DAAD19-02-1-0040.\\
\\

\end{document}